# Strain and twist angle driven electronic structure evolution in twisted bilayer graphene

Jiawei Yu,[1,2†] Guihao Jia,[1,2†] Qian Li,[1,2†] Zhen Zhan,[3†] Yuyang Wang,[1,2] Kebin Xiao,[1,2] Yongkang Ju,[4] Hongyun Zhang,[1,2] Zhiqiang Hu,[1,2] Yunkai Guo,[1,2] Biao Lian,[5] Peizhe Tang,[4] Pierre A. Pantaleón,[3*] Shuyun Zhou,[1,2] Francisco Guinea,[3,6] Qi-Kun Xue,[1,2,7,8,9*] and Wei Li[1,2,9*]

[1]*State Key Laboratory of Low-Dimensional Quantum Physics, Department of Physics, Tsinghua University, Beijing 100084, China*

[2]*Frontier Science Center for Quantum Information, Beijing 100084, China*

[3]*Imdea Nanoscience, Faraday 9, 28015 Madrid, Spain*

[4]*School of Materials Science and Engineering, Beihang University, Beijing 100191, China*

[5]*Department of Physics, Princeton University, Princeton, New Jersey 08544, USA*

[6]*Donostia International Physics Center, Paseo Manuel de Lardizábal 4, San Sebastián 20018, Spain*

[7]*Beijing Academy of Quantum Information Sciences, Beijing 100193, China*

[8]*Southern University of Science and Technology, Shenzhen 518055, China*

[9]*Hefei National Laboratory, Hefei 230088, China*

[*]To whom correspondence should be addressed: pierre.pantaleon@imdea.org; qkxue@mail.tsinghua.edu.cn; weili83@tsinghua.edu.cn

## Abstract

In twisted bilayer graphene (TBG) devices, local strains frequently coexist and intertwine with the twist-angle-dependent moiré superlattice, significantly influencing the electronic properties of TBG, yet their combined effects remain incompletely understood. Here, using low-temperature scanning tunneling microscopy, we study a TBG device exhibiting both a continuous twist-angle gradient from 0.35° to 1.30° and spatially varying strain fields, spanning the first (1.1°), second (0.5°) and third (0.3°) magic angles. We visualize the evolution of flat and remote bands in energy and real space with atomic resolution. Near the first magic angle, we discover an anomalous spectral weight transfer between the two flat band peaks, signifying the role of strain and electronic correlations, as further evidenced by an unusual spatial dispersion of these peaks within a moiré unit cell. In contrast, remote band peak energy offers a strain-insensitive indicator of the local twist angle. Structural analysis further reveals non-negligible shear strain across the sample. All observations are quantitatively reproduced by a continuum model that incorporates heterostrain and a self-consistent Hartree potential, revealing the critical but unexplored role of shear strain in shaping the low-energy electronic landscape of TBG.

Magic-angle twisted bilayer graphene (MATBG) has attracted enormous attention due to its unique electronic structures, which combine strong correlations and intriguing topology[1-27]. When the twist angle approaches the first magic angle ($\theta_{m1}\sim1.1°$), the band structure of TBG is strongly modified by the moiré superlattice formed between the two graphene layers, giving rise to two nearly flat bands (FBs)[1-5]. As the FBs are partially filled, the greatly enhanced electronic interactions can give rise to a variety of novel phenomena, such as correlated insulating phases[4-9], unconventional superconductivity[10-14], quantum anomalous Hall states[15], and transitions between multiple topological phases[16-21]. Moreover, multiple flat bands have been observed near the second magic angle ($\theta_{m2}\sim0.5°$)[28].

Local strains[5,29-37] and twist angle disorder[3,38,39] may form during device fabrication. Strain can cause the broadening of the FBs and affect the electronic correlation[30,33], and it may also induce inhomogeneity in the twist angle[32]. The twist angle gradient can generate intrinsic band bending[38]. As a consequence, the electronic properties may vary significantly even across devices with similar nominal twist angles[33-35]. However, a systematic understanding of how local strain and twist angle govern the electronic properties of TBG remains incomplete. Interestingly, such device imperfections can be turned into an advantage: a device exhibiting continuous spatial variations in both twist angle and local strain offers an ideal platform for comprehensively investigating their individual and combined effects on the electronic properties of TBG[32].

In this study, we use scanning tunneling microscopy/spectroscopy (STM/STS) to investigate a specially fabricated TBG device featuring a continuous gradient of twist angles from 0.35° to 1.30° along with substantial strain variations. This range spans the first ($\theta_{m1}$, 1.1°), second ($\theta_{m2}$, 0.5°), and third ($\theta_{m3}$, 0.3°) magic angles, enabling us to directly map the evolution of the TBG electronic structure as a function of both twist angle and local strain within a single device.

**Structural Characterization**

Our TBG sample is fabricated on a hexagonal boron nitride (hBN) substrate (device schematic in Fig. 1a). The STM topography (Fig. 1b) shows an extensive flat area with a periodic array of bright spots forming a deformed triangular moiré pattern, outlined by red dashed lines. As illustrated in the lower-right panel of Fig. 1b, the moiré pattern comprises three distinct regions: AA sites (bright spots), AB/BA domain, and domain walls (DWs) separating AB and BA domains[34].

Across the sample area (see the full image in Supplementary Sec. S1), the spacing between AA sites (moiré wavelength) varies smoothly, indicating a continuous spatial evolution of the twist angle. The systematic rotation and distortion of moiré triangles reveal the presence of a strain field, which comprises three components: $\varepsilon_{xx}$, $\varepsilon_{yy}$ and $\varepsilon_{xy}$ (Fig. 1c)[33,36].

To precisely analyze the twist angle and strain, we focus on the area within the yellow dashed box in Fig. 1b (full area in Supplementary Sec. S1). This area is displayed in the left panel of Fig. 1d, with all AA sites numbered. We then measure the three moiré wavelengths ($L_1$, $L_2$ and $L_3$) of each moiré triangle, as defined by the orange triangle in the left panel of Fig. 1d. The extracted values vary smoothly and continuously across the area and interchange in specific regions.

Using a geometric model that incorporates the elastic energy of the moiré triangle (Supplementary Sec. S2 and S3), we extract the local twist angle and strain tensor for each AA site. The twist angle varies continuously from 0.35° to 1.30° across the area (right $y$-axis in Fig. 1d), accompanied by significant spatial variations in strain components.

Previous studies have often considered strain in TBG as uniaxial heterostrain with a diagonalized tensor satisfying the Poisson relation[5,7,33,36]. In our device, however, the extracted values indicate that shear strain ($\varepsilon_{xy}$ of a traceless strain tensor) is non-negligible over a large portion of the sample (Supplementary Sec. S3A). This is consistent with the fact that a traceless strain tensor minimizes the elastic energy, making such configurations energetically favorable.

Strictly, based on a detailed analysis of the strain tensor components (Supplementary Sec. S3B), the area within the yellow box (Fig. 1d) is divided into three regions: region-I (red shading, AA01-AA10, 0.35°-0.58°) with approximately uniaxial strain; region-II (white, AA11-AA22, 0.58°-0.98°) with nearly pure shear strain (traceless); region-III (blue shading, AA23-AA60, 0.98°-1.30°) with mixed strain components.

This analysis confirms that our sample hosts a spatially varying twist angle and strain field, providing a well-defined platform to study their effects on the local electronic structure. As demonstrated in the following sections, these twist angles and strain components, particularly shear strain, directly influence the local electronic properties revealed by spectroscopy.

**Tunneling Spectra at Magic Angles**

We then investigate how the moiré geometry modifies the electronic structure within a single device. Specifically, we access areas with twist angles near $\theta_{m1}$, $\theta_{m2}$ and $\theta_{m3}$, enabling a systematic comparison of their spectroscopic signatures. We focus on tunneling spectra (d$I$/d$V$ spectra) acquired at selected AA sites (Fig. 1e; locations marked by red dots in Fig. 1d). All spectra exhibit pronounced features (higher density of states) between -90 meV and -30 meV, consistent with narrow moiré bands. In every case, the Fermi level ($E_F$) lies above this energy window, indicating fully filled FBs and intrinsic electron doping of the device.

Spectra at AA33, AA45, and AA53 sites display the characteristic double-peak features of two FBs[5] near $\theta_{m1}$ in region-III. Although moderate strain broadens the peaks[30,33], the two FBs remain spectroscopically distinguishable. By contrast, spectra at AA09 and AA03, corresponding to $\theta_{m2}$ and $\theta_{m3}$ respectively in region-I, exhibit broader peaks. This is consistent with the multiple moiré bands picture at higher-order magic angles[23,28]. Additionally, strain, combined with intrinsic doping, enlarges the bandwidth and leads spectral overlap with nearby states. Consequently, spectra near $\theta_{m2}$ and $\theta_{m3}$ exhibit a diffuse character.

**Spatial Evolution of Energy Bands Across the Moiré Triangle**

Figure 1f depicts the d$I$/d$V$ colormap acquired along the orange dashed line connecting AA32 to AA29 in Fig. 1d within region-III, capturing the spatial evolution of the local density of states (LDOS) of electrons across the route of AA-BA-DW-AB-AA near $\theta_{m1}$. In the FB energy range, the LDOS is maximized at AA sites and suppressed in the AB/BA domains and at the DWs. Above $E_F$, at the energies of remote conduction bands (RCBs), the LDOS is instead enhanced in the AB/BA domains and DWs compared with AA sites. These observations indicate that the FB states are predominantly localized at AA sites, whereas the RCB states are more delocalized and extend over the AB/BA domains and DWs[8].

Notably, as the STM tip moves from the AA center toward the AB/BA domains, the two FB peaks shift towards $E_F$, indicating an unusual spatial dispersion of the FB across a single moiré triangle. In contrast, the energy of the RCB freezes over AB/BA domains and DWs.

These observations are further confirmed by Fig. 1g, which shows the d$I$/d$V$ colormap acquired along the arrowed white dashed line (Fig. 1d) across a series of AA sites and DWs over the entire range of twist angles in the studied area. An oscillatory evolution of FBs (features below $E_F$) and RCBs (features above $E_F$) is observed, synchronized with the periodicity of alternating AA sites and DWs. This indicates that both the spatial redistribution of electronic states within each moiré triangle and the spatial dispersion of FB are ubiquitous throughout the sample and largely independent of the local strain composition. Additionally, a noticeable trend emerges: both the FBs and RCBs peaks shift towards the $E_F$ with decreasing twist angle.

**Strain and Twist Angle Dependent Evolution of Flat Bands**

To further investigate the evolution of the FBs with twist angle and strain, we extract d$I$/d$V$ spectra at multiple AA sites across twist angles from 0.58° to 1.30° (Fig. 2a). The corresponding colormap of Fig. 2a further highlights the evolution of the FBs (Fig. 2b). From 1.30° to 0.80°, the peaks associated with the lower flat band (FB1) and the upper flat band (FB2) are generally well separated. As the twist angle decreases, the FB1 and FB2 peaks converge in energy and abruptly merge into a single peak (FB3) between 0.78° and 0.72° in region-II. This merging coincides with a drop in the pure shear strain component ($\varepsilon_{xy}$) from finite to vanishingly small values (Fig. 2b and Supplementary Sec. S3), consistent with a shear-strain-driven energy separation of the FBs.

More importantly, near $\theta_{m1}$ in region-III, we observe a clear spectral weight transfer from FB1 to FB2, resulting in an extremely sharp FB2 peak and a weakened FB1 peak (see white dashed box in Fig. 2b). This spectral weight transfer is anomalous, as the FBs are theoretically predicted to be nearly symmetric with comparable spectral weight[30,32], suggesting the influence of additional factors (such as strain or electronic interactions) that may break this symmetry.

To quantify the anomalous spectral weight transfer, we extract the peak positions (Fig. 2c), heights (Fig. 2d), and widths (Fig. 2e) of the FBs from the d$I$/d$V$ spectra (Supplementary Sec. S4). As the twist angle decreases from 1.30° to 1.20°, the FB2 peak shifts slightly toward $E_F$, then maintains an almost fixed energy from 1.20° to 0.78° until it becomes indistinguishable. In contrast, the FB1 peak shifts toward $E_F$ but exhibits pronounced energy fluctuations over the same 1.20° to 0.78° range (Fig. 2c). Furthermore, the energy separation between the two FB peaks remain pronounced as the twist angle changes from 1.30° to 0.78°, a range where the corresponding strain components maintain finite values. Notably, from 0.90° to 0.78° in region-II, where the pure shear strain is finite, the energy separation persists. This separation collapses when the twist angle decreases to 0.72°, where the pure shear strain abruptly drops to zero. This indicates that pure shear strain contributes significantly to the energy separation of FBs.

The anomalous spectral weight transfer from FB1 to FB2 is further confirmed. Near $\theta_{m1}$, the FB2 peak reaches a maximum height (Fig. 2d) and a minimum full width at half maximum (FWHM; Fig. 2e), representing the "flattening" behavior of the moiré band[40]. In contrast, the FB1 peak shows the opposite behavior and is almost invisible. When the twist angle decreases to 1.05° (see the intersection of the guide lines in Fig. 2d and e), the height of FB2 starts to decrease while FB1 gradually recovers. Previous studies on partially filled FBs have shown that spectral weight redistribution of the FB peaks[8] and the "flattening" evolution of the FB with twist angle[25] may correspond to the strongly correlated electronic interactions. Our results demonstrate that, even at the full filling of the FBs, the interaction effects between the two FBs also need to be reconsidered, as they may be significantly modified by both twist angle and strain. Indeed, these trends are consistent with the combined effect of mixed strain and full filling of FBs due to intrinsic electron doping, and are reproduced by our subsequent calculations.

Figure 2f shows STM topography and LDOS maps of an AA site at a twist angle of 1.05°. The spatial distributions of both FB1 (-73 meV) and FB2 (-46 meV) are concentrated at the AA site center. FB2 exhibits stronger localization, while FB1 shows broader spatial extension. At energies away from the FB energies, the elongated LDOS spot features at the AA site gradually spread outward from the center, forming ring-like features[41]. These rings are non-circular and elongated, reflecting strain-induced anisotropy[32]. The spatial anisotropy is more pronounced for FB1 than for FB2 (as indicated by the double-headed arrows in Fig. 2f; see Supplementary Sec. S5). The irregular shapes of the FB spatial distributions provide additional evidence for the presence of strain in the studied area.

**Twist Angle Dependent Evolution of Remote Conduction Bands**

The remote conduction band peaks (RCBPs) are predominantly observed in AB/BA domains and DWs, where the FB contribution to the LDOS is not dominant. The d$I$/d$V$ spectra acquired at the centers of AB/BA domains and DWs track the evolution of the first (RCBP1) and the second (RCBP2) remote conduction band peaks as a function of twist angle (Fig. 3a-d). The twist angle of a DW is determined by averaging the twist angles of two connected AA sites, whereas for an AB/BA domain, it is defined as the average of the three vertex AA sites.

The RCBP energies exhibit a clear one-to-one relationship with twist angle (Fig. 3e) and remain largely insensitive to strain (Supplementary Sec. S6). As shown in Fig. 1f, the RCBP energies remain fixed across AB/BA domains and DWs within a moiré triangle, making them reliable indicators of local twist angle throughout the sample. Practically, by selecting specific RCBP1 energies in LDOS maps, we can identify regions sharing identical twist angle (Fig. 3f; Supplementary Sec. S7). From these highlighted areas, the direction of the twist angle gradient across the sample can be easily determined. These observations are consistent with recent reports establishing remote bands as robust fingerprints of the local twist angle in TBG[42].

**Theory**

In the preceding sections, we analyzed the geometric structure of the studied area and described the corresponding evolution of the electronic structures in both energy and real space, identifying the important roles of twist angle and strain. To enable direct comparison, Fig. 4 presents theoretical calculations to mirror the experimental geometric and electronic structures, using a low-energy continuum Hamiltonian for TBG[1,43] that incorporates heterostrain[32,44] and electrostatic interactions at the Hartree level[45] (Supplementary Sec. S8).

We first calculate the low-energy band structures at $\theta_{m1}$. Under full filling of the FBs, the band structures calculated without (Fig. 4a) and with (Fig. 4b) considering the mixed strain components ($\varepsilon_{xx}+\varepsilon_{yy}$ ~ -0.5%, $\varepsilon_{xy}$ ~ -0.8%) show distinct characteristics, with all bands labeled. Specifically, FB1 and FB2 are close in energy without strain, whereas they separate significantly when strain is included. Additionally, the energy range of the FBs differs between the two cases, and the energy separations between FBs and remote bands are large enough for distinguishing the well-isolated FB peaks. These calculations confirm that strain reshapes the low-energy bands, drives the energy separation, and increases the bandwidths of the FBs.

We then calculate the LDOS evolution as a function of twist angle (Fig. 4c and d) under the same strain composition as in the studied area (Supplementary Sec. S9 and S10), paralleling the experimental trends shown in Fig. 1e and 2b. For the FBs, the energy separation from 0.78° to 1.30° is reproduced, and the spectral weight transfer between the two FBs near $\theta_{m1}$ is captured in terms of its trend.

Our calculations further demonstrate that both electronic interactions and mixed strain govern the spectral weight transfer of the FBs and their mutual separation. Non-interacting models with the same strain composition predict nearly symmetric FBs with comparable spectral weight (Supplementary Sec. S10), whereas intrinsic electron doping generates a Hartree potential that reshapes the bands and sharpens FB2 (Supplementary Sec. S11). The additional broadening of FB1 and energy separation of FBs observed experimentally are consistent with the presence of mixed strain, as neither pure shear nor pure uniaxial strain alone can reproduce the observed energy separation and spectral weight transfer of FBs (Supplementary Sec. S11). Moreover, some subtle discrepancies in the spectral weight transfer from 0.84° to 1.06° are noted between calculation and experiment, where the calculated FB1 peak is higher than the FB2 peak (spectrum at 1.06° in Fig. 4c), and the energy separation between FB2 and RCBP1 is significantly smaller in the calculation. When the Fock exchange interaction is included in calculations for twist angle smaller than 1.1°, the spectral weight transfer (lower FB1 and higher FB2 peaks) and the larger energy separation between FB2 and RCBP1 are better reproduced near $\theta_{m1}$ (Supplementary Sec. S12). Additionally, relaxation effects can also contribute to the spectral weight transfer behavior.

Next, we focus on the spatial evolution at $\theta_{m1}$. Figure 4e presents the calculated LDOS along the path defined in Fig. 1f, capturing both the enhanced RCBP intensity in AB/BA and DW regions and the enhanced FB peaks at AA sites observed experimentally. In our calculations, strain is treated as local variation within the moiré triangle to modify the low-energy model. This local strain variation[46,47], combined with local relaxation[48,49] and exchange interactions[50], contributes to the local variation of Fermi velocity within the moiré triangle (Supplementary Sec. S13), thereby driving the spatial dispersion of the FB peaks. Although the path connects areas with nearly identical local twist angles, the calculated LDOS evolution reveals that strain variation across the moiré triangle, manifesting as a larger FB–RCB gap at AA sites that decreases toward AB/BA domains and DWs, a trend observable in Fig. 1f. We note that this methodology provides a reasonable approximation of the experimental results; a fully exact analysis would require a complete tight-binding model spanning multiple unit cells, which is beyond the scope of this work.

Finally, the monotonic evolution of the RCBPs with twist angle is also reproduced at AB/BA domains and DWs, as shown in Fig. 4f and Supplementary Sec. S9. Notably, our calculations in Supplementary Sec. S10-S11 show that the RCBP energy remains insensitive to strain variation at fixed doping levels, consistent with the experimental observations.

**Discussion**

Our study establishes that the interplay between local strain and twist angle fundamentally reshapes the electronic structure of TBG. The sample exhibits three distinct strain regions, with region-II dominated by pure shear strain, where the traceless tensor minimizes elastic energy. Crucially, we observe two key phenomena near $\theta_{m1}$: a pronounced spectral weight transfer between the flat bands (Fig. 2a, b) and a shear-strain-sustained energy separation that collapses when the shear strain vanishes. These features are quantitatively captured by our model incorporating heterostrain and Hartree potential (Fig. 4c, d), demonstrating that both electrostatic interactions and non-uniform strain, particularly shear, are essential for breaking band symmetry. Furthermore, strain variation within moiré unit cells drives spatial dispersion of flat bands (Fig. 1f), while remote bands provide a strain-insensitive probe of twist angle (Fig. 3e, f). Our findings underscore that engineering correlated states in moiré materials requires concurrent control of lattice deformation and electron interactions.

## Methods

### STM measurements

Our experiments were carried out on an ultrahigh-vacuum commercial STM system (Unisoku) that operated on a constant base temperature of 4.2 K with optical microscope observation. The base pressure of the system is $1.0\times10^{-10}$ Torr. The TBG devices were degassed at 170 °C under ultra-high vacuum before transferring into the STM. The STS data were obtained by a standard lock-in method that applied an additional small a.c. voltage with a frequency of 973 Hz. The $dI/dV$ spectra were collected by disrupting the feedback loop and sweeping the d.c. bias voltage.

### TBG device

The TBG samples were prepared by using the clean dry transfer method. First, monolayer graphene was exfoliated onto a clean $SiO_2$/Si substrate. A BN flake attached to Polydimethylsiloxane (PDMS) stamp was then positioned above the monolayer graphene under an optical microscope. Half of the monolayer graphene was picked up by the BN on PDMS, forming PDMS/BN/Graphene structure. Then the other half of the graphene was rotated to the desired twist angle and picked up by the PDMS/BN/Graphene heterostructure. Subsequently, the PDMS/BN/TBG structure was flipped over, and the BN/TBG sample was picked up by another PDMS stamp and then transferred onto a gold-coated substrate. The cleanness and quality of the TBG sample are critical for obtaining high-quality STM data. Finally, one piece of graphite was used to connect the graphene and the gold-coated substrate to ensure good electrical conductivity during STM measurements.

### Structural analysis

We locate AA sites in the STM topographies and use the triangles formed by three neighboring AA centers to infer the local moiré lattice. From each triangle, we extract three moiré lattice vectors (magnitudes and orientations) and fit them with a geometric model of twisted and strained bilayers to obtain the local twist angle and the in-plane strain tensor components ($\varepsilon_{xx}$, $\varepsilon_{yy}$, $\varepsilon_{xy}$). The problem is over-determined because multiple triangle choices can yield similar vector sets. We select a unique solution by minimizing the elastic energy of the strain field.

### Low-energy continuum model with strain and Hartree interactions

We use a low-energy continuum Hamiltonian for TBG, including heterostrain and electrostatic interactions at the Hartree level. For each locally extracted twist angle and strain tensor (from the structural analysis) we calculate the twist angle dependent LDOS. To obtain the spatial variation of LDOS within the moiré unit cell, we compute a position-resolved LDOS on a grid of points following the paths of the experimental line scans.

Electron-electron interactions are treated self-consistently via a moiré-periodic Hartree potential with a dielectric constant chosen to account for hBN/graphite screening. The total self-consistent Hamiltonian is iterated until the charge and bands converge, with the total filling set by the experimentally inferred electron doping (middle bands fully occupied). The LDOS is evaluated from the converged eigenstates and broadened by Gaussian smoothing to match the effects of the experiment temperature and the STM lock-in modulation, yielding the maps and spectra reported in the main text.

**Data availability**

The data that support the findings of this study are included in this article and its supplementary information file and are available from the corresponding author upon reasonable request.

**Acknowledgements**

J. Yu, G. Jia, Q. Li and Z. Zhan contributed equally to this work. We thank Z. Song and S. Yuan for helpful discussions. We acknowledge Emma Martínez (IES Fortuny) for her help at IMDEA Nanoscience in the analysis of twist-angle data, carried out as part of her participation in the 4ESO+Empresa educational program of the Comunidad de Madrid. The research was supported by the National Science Foundation of China (Grants No. 92365201, No. 51788104, No. 11427903, and No. 12234011) and the Ministry of Science and Technology of China (Grant No. 2022YFA1403100 and No. 2021YFA1400100), the Initiative Research Projects of Tsinghua University (No. 20211080075), and the Beijing Advanced Innovation Center for Future Chip (ICFC). B. L. is supported by the National Science Foundation under award DMR-2141966, and the National Science Foundation through Princeton University's Materials Research Science and Engineering Center DMR-2011750. IMDEA Nanociencia acknowledges support from the 'Severo Ochoa' Programme for Centres of Excellence in R&D (CEX2020-001039-S/AEI/10.13039/501100011033). P.A.P and F.G acknowledges support from NOVMOMAT, project PID2022-142162NB-I00 funded by MICIU/AEI/10.13039/501100011033 and by FEDER, UE as well as financial support through the (MAD2D-CM)-MRR MATERIALES AVANZADOS-IMDEA-NC. F.G acknowledges the support from the Department of Education of the Basque Government through the project No. PIBA\2023\1\0007(STRAINER). Z.Z acknowledges support from the European Union's Horizon 2020 research and innovation programme under the Marie-Sklodowska Curie grant agreement No. 101034431.

**Author contributions**

W.L., S.Z., F.G. and Q.-K.X. conceived and supervised the research project. J.Y., K.X., Z.H. and Y.G. performed the STM experiments. Q.L., H.Z. and S.Z. prepared the TBG device. W.L., J.Y., G.J., Z.Z., Y.W., Y.J., B.L., P.T. and P.A.P. analyzed the data. P.A.P. and Z.Z. provided the theoretical calculations. W.L., J.Y., G.J., Z.Z., Y.W. and P.A.P. wrote the manuscript with input from all other authors.

**Competing interests**

The authors declare no competing interests.

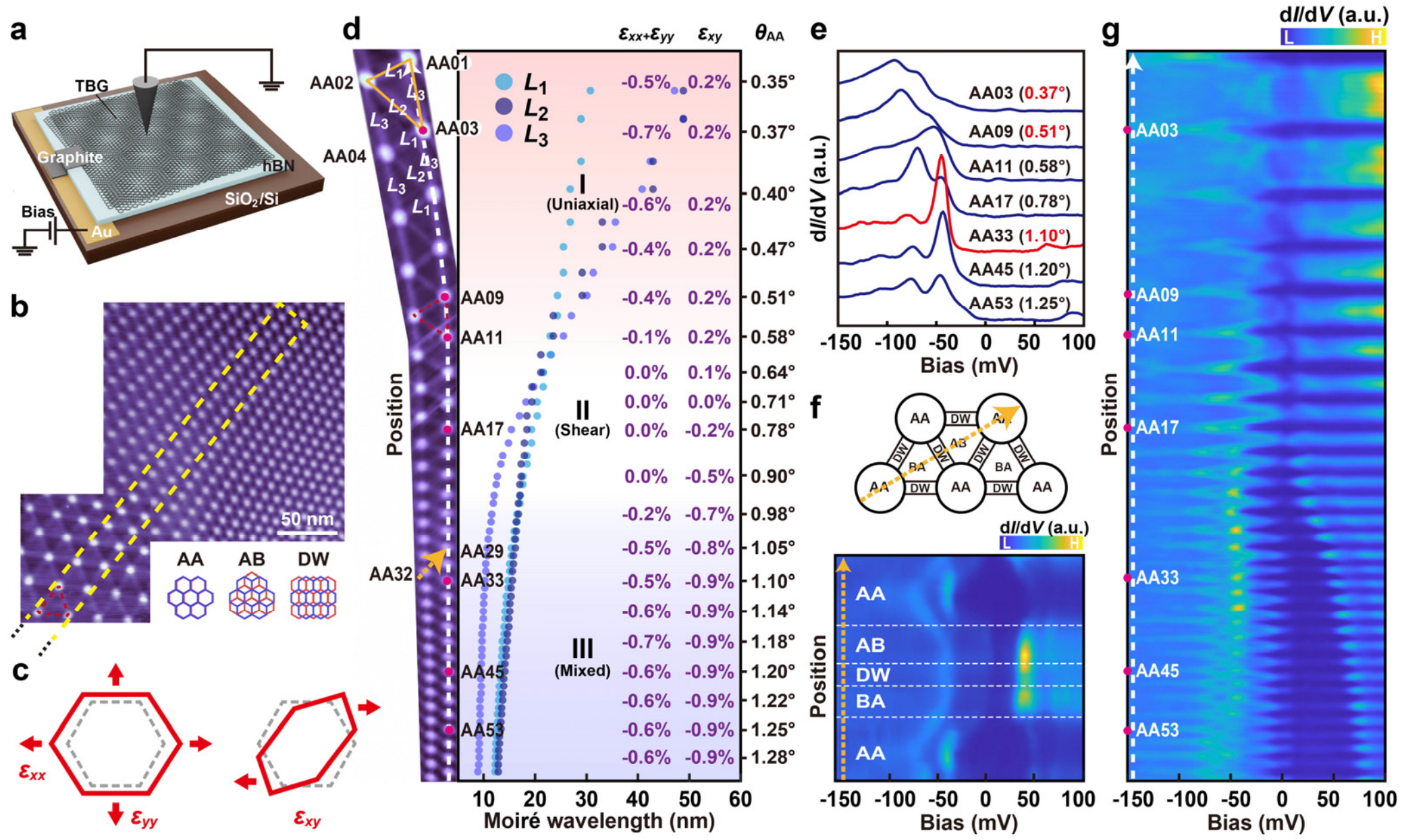

**Fig. 1 | Evolution of tunneling spectra in a strained twisted bilayer graphene device with continuously varying twist angles spanning multiple magic angles. a**, Sample schematic. Bias voltage ($V_b$) is applied between the STM tip and TBG through a graphite electrode. **b**, STM topography of a large area on TBG, a composite of two images (200 nm × 200 nm and 100 nm × 100 nm, $V_b$ = -800 mV, tunneling current $I_t$ = 20 pA), the unclosed yellow dashed box marks the studied area, and the black dots indicate additional measured area. **c**, Schematic of the strain components. Positive values indicate tension and negative values indicate compression. **d**, Moiré wavelengths and corresponding calculated twist angles and strains. The left panel shows the area within the yellow dashed box. The red triangles in **b** and **d** correspond to the same location. $L_1$, $L_2$, $L_3$ are plotted in the right panel. The corresponding calculated twist angles and strain values are shown on the right *y*-axis. The area is divided into three regions based on distinct strain components: region-I (red shading), region-II (white) and region-III (blue shading). **e**, Tunneling spectra taken at centers of seven AA sites. Magic angles are highlighted in red. **f,** d$I$/d$V$ colormap taken along the orange dashed line in **d** across AA, AB, DW, BA, and AA sites. The upper panel indicates the route of the dashed line in detail. **g**, d$I$/d$V$ colormap taken along the arrowed white dashed line in **d**; the locations of the seven AA sites are marked. Set point: **e-g**, $V_s$ = -200 mV, $I_t$ = 200 pA.

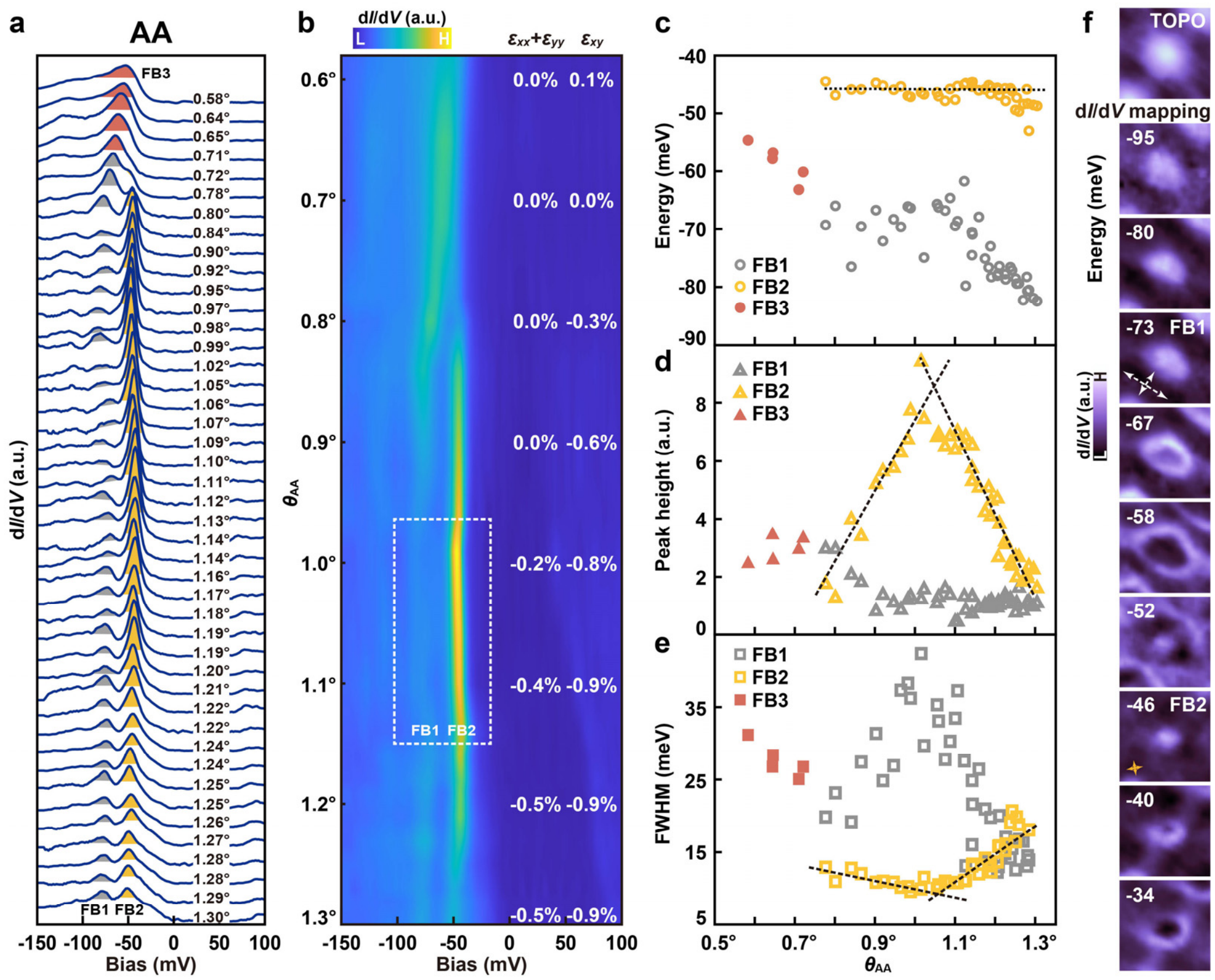

**Fig. 2 | Twist angle dependent evolution of the flat bands at AA sites close to $\theta_{m1}$. a**, Tunneling spectra at AA sites as a function of twist angle $\theta_{AA}$. For clarity, spectra are equally offset. The two well separated FB1 and FB2 peaks are colored gray and yellow, respectively, and the combined FB3 peak is in red. Set point: $V_s$ = -200 mV, $I_t$ = 200 pA. **b**, Colormap of data in **a** to visualize the peak evolution. White dashed box indicates the 0.97°-1.15° range in region-III. **c-e**, Energy, peak height and FWHM of the FB states as a function of $\theta_{AA}$. Black dashed lines indicate the evolution trends. **f**, Topography and d$I$/d$V$ map near the AA29 site (1.05°, 15.4 nm × 15.4 nm). The anisotropy directions of FB1 and FB2 are indicated by gray and yellow double-headed arrows. Set point: $V_s$ = -200 mV, $I_t$ = 200 pA.

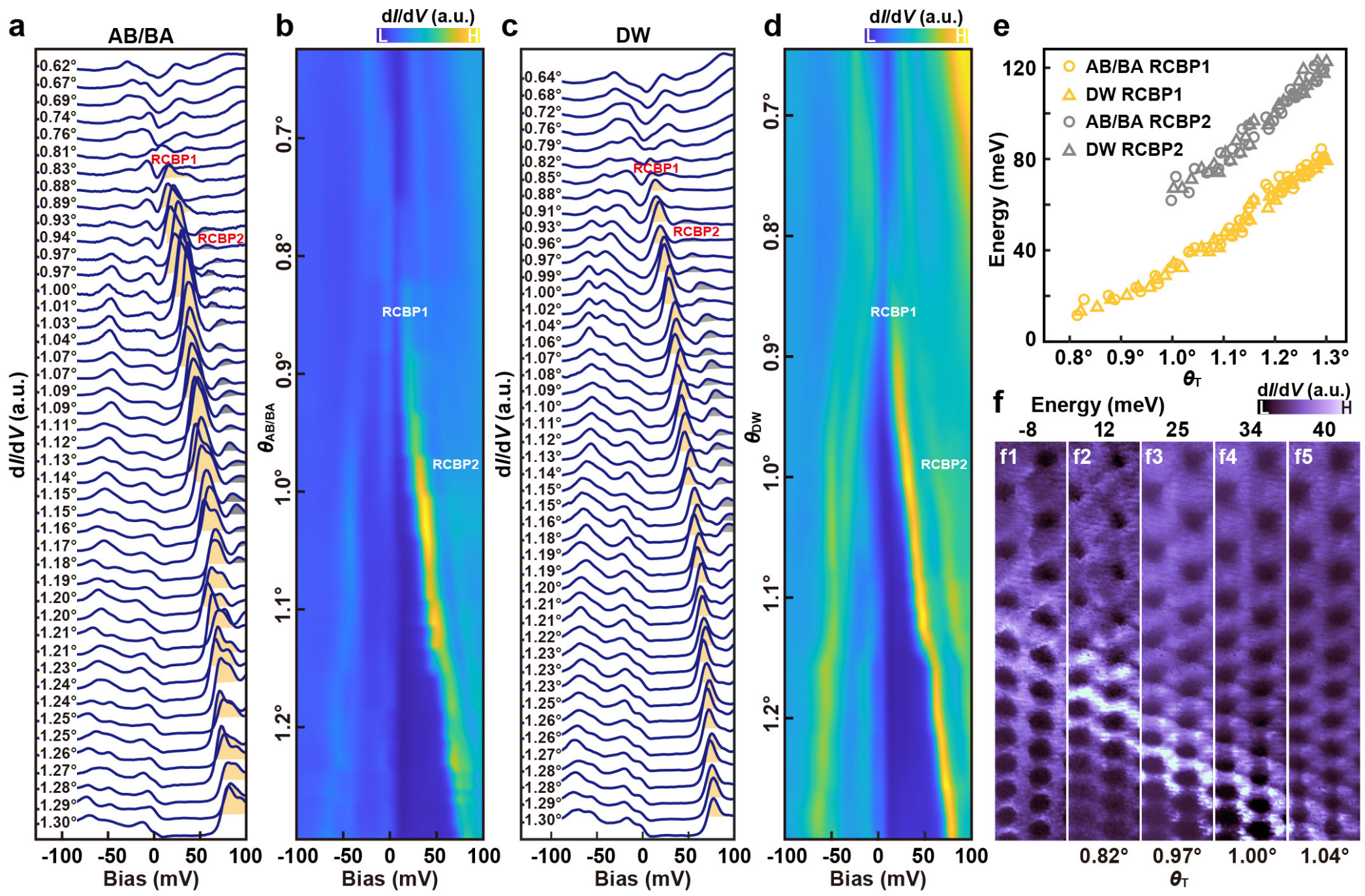

**Fig. 3 | Twist angle dependent evolution of the remote conduction band peaks on AB/BA domains and DWs. a,** Tunneling spectra acquired at AB/BA domain centers as a function of twist angle $\theta_{AB/BA}$. **b**, Corresponding colormap of **a**. **c**, Tunneling spectra on DWs as a function of twist angle $\theta_{DW}$. The RCBP1 and RCBP2 are in yellow and gray, respectively. **d**, Colormap of **c**. **e**, Evolution of RCBP1 and RCBP2 on DWs and AB/BA domains with twist angle $\theta_T$ ($\theta_{AB/BA}$ or $\theta_{DW}$). **f1-f5**, d*I*/d*V* maps acquired at different RCBP1 energies (32 nm × 170 nm). The brighter region in each map corresponds to the spatial distribution of the RCBP1 at a specific energy, with the average twist angles for these brighter regions indicated below. Set point: **a**, **c**, **f**, $V_s$ = -200 mV, $I_t$ = 200 pA.

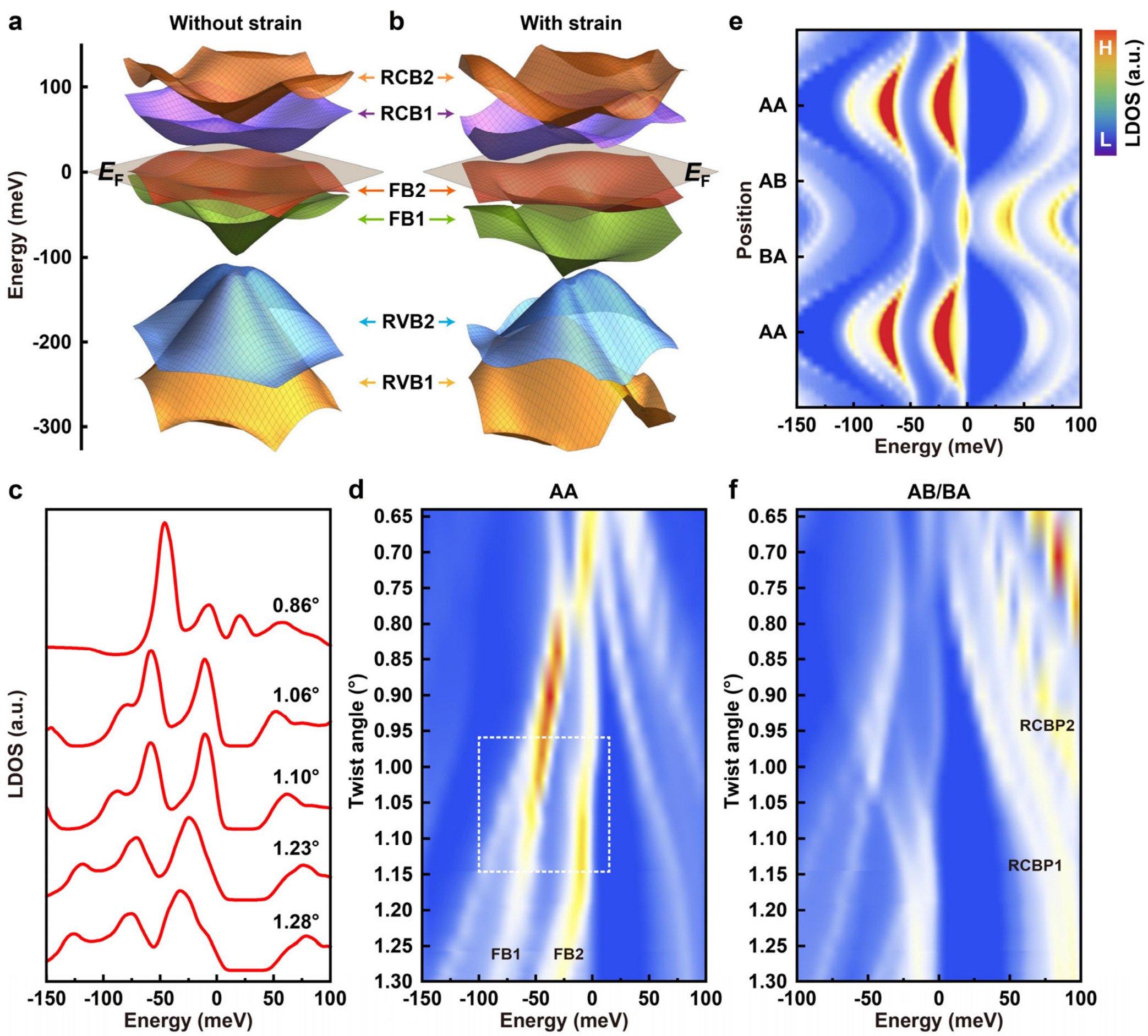

**Fig. 4 | Theoretical evolution of flat and remote conduction band peaks. a, b**, Low-energy band structures calculated at $\theta_{m1}$ under full filling of the FBs, without (**a**) and with (**b**) strain, respectively. Bands are labeled from bottom to top: lower remote valence band (RVB1), upper remote valence band (RVB2), FB1, FB2, RCB1 and RCB2. **c**, Calculated LDOS at AA sites for different twist angles. The calculation incorporates the Hartree potential and the mixed strain profile of the experimental region. The spectra are broadened by Gaussian smoothing to match the experimental conditions. **d**, Calculated LDOS colormap as a function of twist angle at AA sites. The white dashed box marks the same twist angle range as in Fig. 2b. **e**, Calculated LDOS along the path defined in Fig. 1f, capturing the spatial redistribution of FB peaks and RCBPs and the spatial dispersion of the FBs across the moiré unit cell. **f**, Calculated LDOS colormap as a function of twist angle at AB/BA domains, reproducing the monotonic evolution of RCBPs.